\documentclass[preprint2]{emulateapj}
\usepackage{color}
\usepackage{amsmath}
\usepackage{threeparttable}
\usepackage{natbib}
\usepackage{times}
\usepackage{graphicx}
\usepackage[mediumspace,Gray,squaren]{SIunits} 
\newcommand{\rmicron}{$\,\micro$m}

\DeclareGraphicsExtensions{.jpg,.pdf,.png,.eps,.ps}

\newcommand{\spire}{SPIRE\, }
\newcommand{\Herschel}{{\em Herschel}\,}

\def\muK{\rm $\mu${\mbox{K}}}

\def\oz#1{{}}

\def\ell{l}

\newcommand{\comment}[1]{{}}

\def\McGill{1}
\def\Caltech{2}
\def\BCCP{3}
\def\UChicago{4}
\def\KICPChicago{5}
\def\EFIChicago{6}
\def\ArgonneHEP{7}
\def\PhysicsUChicago{8}
\def\JPL{9}
\def\Miss{10}
\def\AAUChicago{11}
\def\Argonne{12}
\def\NIST{13}
\def\ColoradoC{14}
\def\Berkeley{15}
\def\ColoradoH{16}
\def\Davis{17}
\def\LBNL{18}
\def\UBC{19}
\def\Arizona{20}
\def\Michigan{21}
\def\Munich{22}
\def\ExcellenceCluster{23}
\def\MPE{24}
\def\CaseWestern{25}
\def\Minnesota{26}
\def\ArtInstChicago{27}
\def\IPAC{28}
\def\CfA{29}

\begin{document}

\title{A CMB lensing mass map and its correlation with the cosmic infrared background}

\slugcomment{}

\author{
G.~P.~Holder\altaffilmark{\McGill},
M.~P.~Viero\altaffilmark{\Caltech},
O.~Zahn\altaffilmark{\BCCP}, 
K.~A.~Aird\altaffilmark{\UChicago},
B.~A.~Benson\altaffilmark{\KICPChicago,\EFIChicago},
S.~Bhattacharya\altaffilmark{\KICPChicago,\ArgonneHEP},
L.~E.~Bleem\altaffilmark{\KICPChicago,\PhysicsUChicago},
J.~Bock\altaffilmark{\Caltech, \JPL},
M.~Brodwin\altaffilmark{\Miss},
J.~E.~Carlstrom\altaffilmark{\KICPChicago,\EFIChicago,\PhysicsUChicago,\AAUChicago,\Argonne},
C.~L.~Chang\altaffilmark{\KICPChicago,\EFIChicago,\Argonne},
H-M.~Cho\altaffilmark{\NIST},
A.~Conley\altaffilmark{\ColoradoC},
T.~M.~Crawford\altaffilmark{\KICPChicago,\AAUChicago},
A.~T.~Crites\altaffilmark{\KICPChicago,\AAUChicago},
T.~de~Haan\altaffilmark{\McGill},
M.~A.~Dobbs\altaffilmark{\McGill},
J.~Dudley\altaffilmark{\McGill},
E.~M.~George\altaffilmark{\Berkeley},
N.~W.~Halverson\altaffilmark{\ColoradoH},
W.~L.~Holzapfel\altaffilmark{\Berkeley},
S.~Hoover\altaffilmark{\KICPChicago,\PhysicsUChicago},
Z.~Hou\altaffilmark{\Davis},
J.~D.~Hrubes\altaffilmark{\UChicago},
R.~Keisler\altaffilmark{\KICPChicago,\PhysicsUChicago},
L.~Knox\altaffilmark{\Davis},
A.~T.~Lee\altaffilmark{\Berkeley,\LBNL},
E.~M.~Leitch\altaffilmark{\KICPChicago,\AAUChicago},
M.~Lueker\altaffilmark{\Caltech},
D.~Luong-Van\altaffilmark{\UChicago},
G.~Marsden\altaffilmark{\UBC},
D.P.~Marrone\altaffilmark{\Arizona},
J.~J.~McMahon\altaffilmark{\Michigan},
J.~Mehl\altaffilmark{\KICPChicago,\Argonne},
S.~S.~Meyer\altaffilmark{\KICPChicago,\EFIChicago,\PhysicsUChicago,\AAUChicago},
M.~Millea\altaffilmark{\Davis},
J.~J.~Mohr\altaffilmark{\Munich,\ExcellenceCluster,\MPE},
T.~E.~Montroy\altaffilmark{\CaseWestern},
S.~Padin\altaffilmark{\KICPChicago,\AAUChicago,\Caltech},
T.~Plagge\altaffilmark{\KICPChicago,\AAUChicago},
C.~Pryke\altaffilmark{\Minnesota},
C.~L.~Reichardt\altaffilmark{\Berkeley},
J.~E.~Ruhl\altaffilmark{\CaseWestern},
J.~T.~Sayre\altaffilmark{\CaseWestern},
K.~K.~Schaffer\altaffilmark{\KICPChicago,\EFIChicago,\ArtInstChicago},
B.~Schulz\altaffilmark{\Caltech, \IPAC},
L.~Shaw\altaffilmark{\McGill},
E.~Shirokoff\altaffilmark{\Berkeley},
H.~G.~Spieler\altaffilmark{\LBNL},
Z.~Staniszewski\altaffilmark{\CaseWestern},
A.~A.~Stark\altaffilmark{\CfA},
K.~T.~Story\altaffilmark{\KICPChicago,\PhysicsUChicago},
A.~van~Engelen\altaffilmark{\McGill},
K.~Vanderlinde\altaffilmark{\McGill},
J.~D.~Vieira\altaffilmark{\Caltech}, 
R.~Williamson\altaffilmark{\KICPChicago,\AAUChicago}, and
M.~Zemcov\altaffilmark{\Caltech,\JPL}
}

\altaffiltext{\McGill}{Department of Physics, McGill University, Montreal, Quebec H3A 2T8, Canada}
\altaffiltext{\Caltech}{California Institute of Technology, Pasadena, CA, USA 91125}
\altaffiltext{\BCCP}{Berkeley Center for Cosmological Physics, Department of Physics, University of California, and Lawrence Berkeley National Laboratory, Berkeley, CA, USA 94720}
\altaffiltext{\UChicago}{University of Chicago, Chicago, IL, USA 60637}
\altaffiltext{\KICPChicago}{Kavli Institute for Cosmological Physics, University of Chicago, Chicago, IL, USA 60637}
\altaffiltext{\EFIChicago}{Enrico Fermi Institute, University of Chicago, Chicago, IL, USA 60637}
\altaffiltext{\ArgonneHEP}{High Energy Physics Division, Argonne National Laboratory, Argonne, IL, USA 60439}
\altaffiltext{\PhysicsUChicago}{Department of Physics, University of Chicago, Chicago, IL, USA 60637}
\altaffiltext{\JPL}{Jet Propulsion Laboratory, Pasadena, CA, USA 91109}
\altaffiltext{\Miss}{Department of Physics and Astronomy, University of Missouri, Kansas City, MO 64110, USA}
\altaffiltext{\AAUChicago}{Department of Astronomy and Astrophysics, University of Chicago, Chicago, IL, USA 60637}
\altaffiltext{\Argonne}{Argonne National Laboratory, Argonne, IL, USA 60439}
\altaffiltext{\NIST}{NIST Quantum Devices Group, Boulder, CO, USA 80305}
\altaffiltext{\ColoradoC}{Center for Astrophysics and Space Astronomy, University of Colorado, Boulder, CO, USA 80309}
\altaffiltext{\Berkeley}{Department of Physics, University of California, Berkeley, CA, USA 94720}
\altaffiltext{\ColoradoH}{Department of Astrophysical and Planetary Sciences and Department of Physics, University of Colorado, Boulder, CO, USA 80309}
\altaffiltext{\Davis}{Department of Physics, University of California, Davis, CA, USA 95616}
\altaffiltext{\LBNL}{Physics Division, Lawrence Berkeley National Laboratory, Berkeley, CA, USA 94720}
\altaffiltext{\UBC}{Department of Physics and Astronomy, University of British Columbia, Vancouver, BC, Canada V6T 1Z1}
\altaffiltext{\Arizona}{Steward Observatory, University of Arizona, Tucson, AZ, USA 85721}
\altaffiltext{\Michigan}{Department of Physics, University of Michigan, Ann  Arbor, MI, USA 48109}
\altaffiltext{\Munich}{Department of Physics, Ludwig-Maximilians-Universit\"{a}t, 81679 M\"{u}nchen, Germany}
\altaffiltext{\ExcellenceCluster}{Excellence Cluster Universe, 85748 Garching, Germany}
\altaffiltext{\MPE}{Max-Planck-Institut f\"{u}r extraterrestrische Physik, 85748 Garching, Germany}
\altaffiltext{\CaseWestern}{Physics Department, Center for Education and Research in Cosmology and Astrophysics, Case Western Reserve U
 niversity,Cleveland, OH, USA 44106}
\altaffiltext{\Minnesota}{Department of Physics, University of Minnesota, Minneapolis, MN, USA 55455}
\altaffiltext{\ArtInstChicago}{Liberal Arts Department, School of the Art Institute of Chicago, Chicago, IL, USA 60603}
\altaffiltext{\IPAC}{Infrared Processing and Analysis Center, California Institute of Technology, JPL, Pasadena, CA USA 91125}
\altaffiltext{\CfA}{Harvard-Smithsonian Center for Astrophysics, Cambridge, MA, USA 02138}

\email{holder@physics.mcgill.ca}

\begin{abstract}

We use a temperature  map of the cosmic microwave background (CMB) obtained
using the South Pole Telescope at 150 GHz to construct
a map of the gravitational convergence to $z\sim 1100$, revealing the fluctuations
in the projected mass density. This map shows 
individual features that are significant at the $\sim 4 \sigma$ level,
providing the first image of CMB lensing convergence.
We cross-correlate this map with \Herschel/\spire maps covering 90 square degrees at
wavelengths of 500, 350, and 250 \rmicron. We show that these
submillimeter-wavelength (submm) maps are strongly correlated with the lensing convergence
map, with detection significances in each of the three submm bands
ranging from 6.7 to 8.8 $\sigma$. We fit the measurement of the
cross power spectrum assuming a simple constant bias model
and infer bias factors
of $b=1.3-1.8$, with a statistical uncertainty of $15\%$,
depending on the assumed 
model for the redshift distribution of the dusty galaxies that are 
contributing to the \Herschel/\spire maps. 

\end{abstract}


\section{Introduction}

Gravitational lensing of the cosmic microwave background (CMB)
is emerging as a powerful cosmological tool.
The spatial variation of the statistical properties of the CMB that is induced
by gravitational lensing was first detected in cross-correlation with radio-selected
galaxy catalogs \citep{smith07, hirata08}, and subsequently detected 
internally in CMB maps by the Atacama Cosmology Telescope collaboration \citep{das11} and the
South Pole Telescope (SPT) collaboration \citep{vanengelen12}. In \citet{bleem12b}, reconstructions of
the mass distribution were found to correlate strongly with galaxy catalogs 
selected in both the optical and infrared bands, while 
\citet{sherwin12} showed that CMB lensing was well-correlated with 
quasars.

Using the CMB as the background source to study gravitational lensing by intervening structure 
offers several advantages over using distant galaxies:
the source redshift is the same for all lines of sight, is extremely well-known,
and has the highest redshift observable with electromagnetic radiation. 
The statistical properties of the source are well-characterized, and 
CMB maps cover areas ranging from a few hundred square degrees
to the full sky. 
However, the single redshift for the CMB does not allow any information about the redshift distribution of the mass along the line of sight, and noise levels in CMB lensing convergence maps at current sensitivities are substantially higher than noise levels in cosmic shear measurements.

As CMB lensing is an integral along the entire line of sight, the strongest cross-correlations will
be with sources that have a similarly broad extent in redshift space. 
As demonstrated below, and as theoretically predicted \citep{song03}, 
the cosmic infrared background (CIB) fluctuations provide an excellent match. The CIB
at submm wavelengths is believed to have a substantial contribution from sources from 
redshifts $z\sim 0.5-3$ \citep{lagache04, amblard11, bethermin11, viero12}.

In this paper, we cross-correlate a map of the gravitational lensing
convergence (proportional to the surface density) derived from SPT
temperature data at 150 GHz with maps of the submillimeter-wavelength (submm) 
sky at 500, 350, and 250 \rmicron\ obtained
with \Herschel/SPIRE. 
By using maps rather than catalogs, as was done in previous CMB lensing cross-correlations,
we are studying emission from sources that are individually unresolved.
The SPT and \Herschel datasets are described in 
sections 2 and 3, and the results of the cross-correlation are presented
in section 4. A comparison with a simple theoretical model is
presented in section 5, and we conclude with a discussion of the results.

\section{CMB Map and Corresponding Mass Map}

The SPT has been used to image 2500 square degrees to a depth of 
$\lesssim$18 \muK-arcmin at 150 GHz, 
and two $\sim 100$-square-degree fields (each subtending  
1h in Right Ascension and 10 degrees in Declination) within this area to 
a depth of $\sim$13 \muK-arcmin. For this work, we use observations 
centered on one of those deeper fields, centered at (RA,DEC)=(23h30m, -55d00m),
using data from both the 2008 and 2010 observing seasons; 
the recent CMB power spectrum measurements of 
\citet{story12} used only the data from 2008 for this field. 

\begin{figure*}
\plottwo{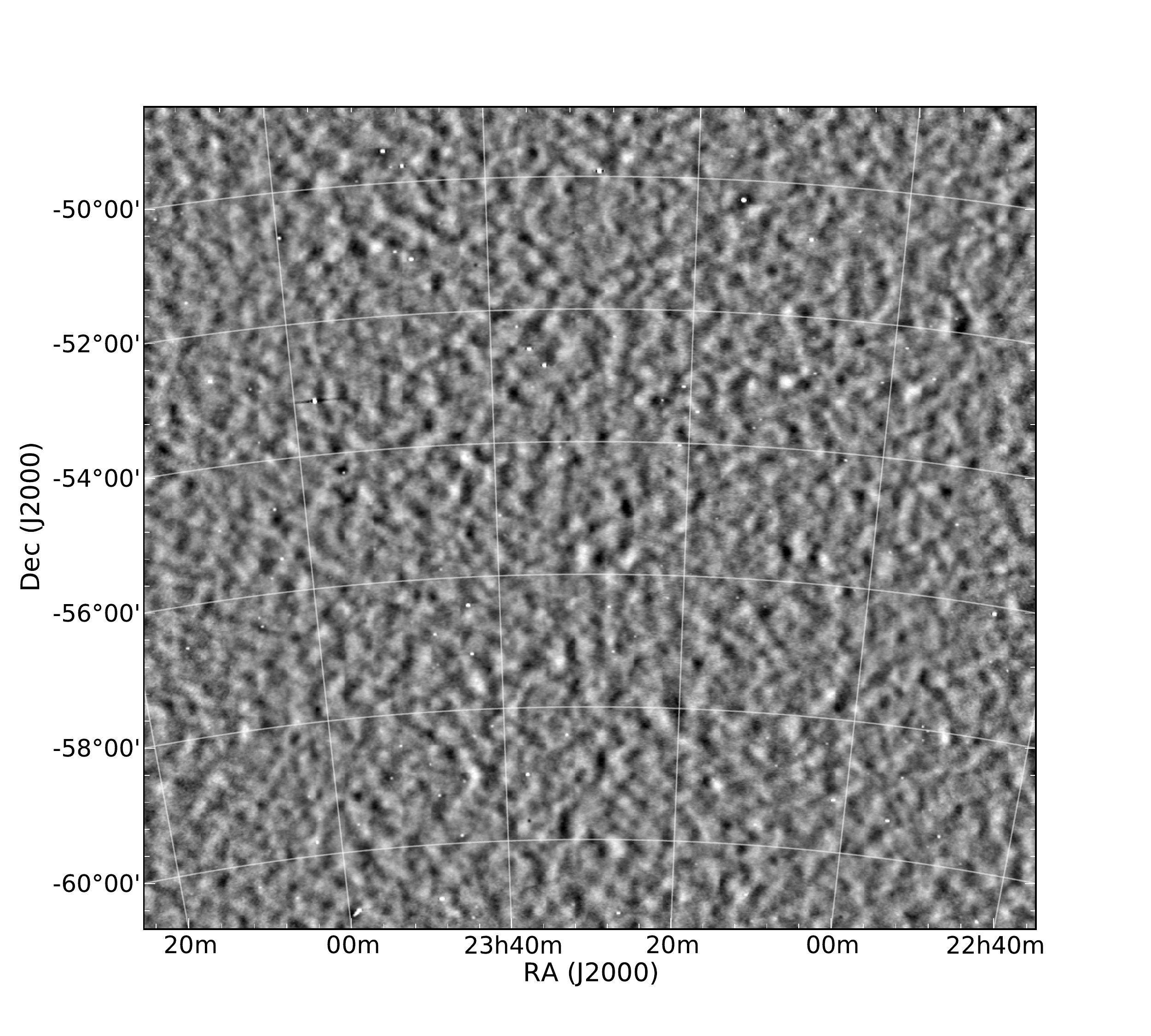}{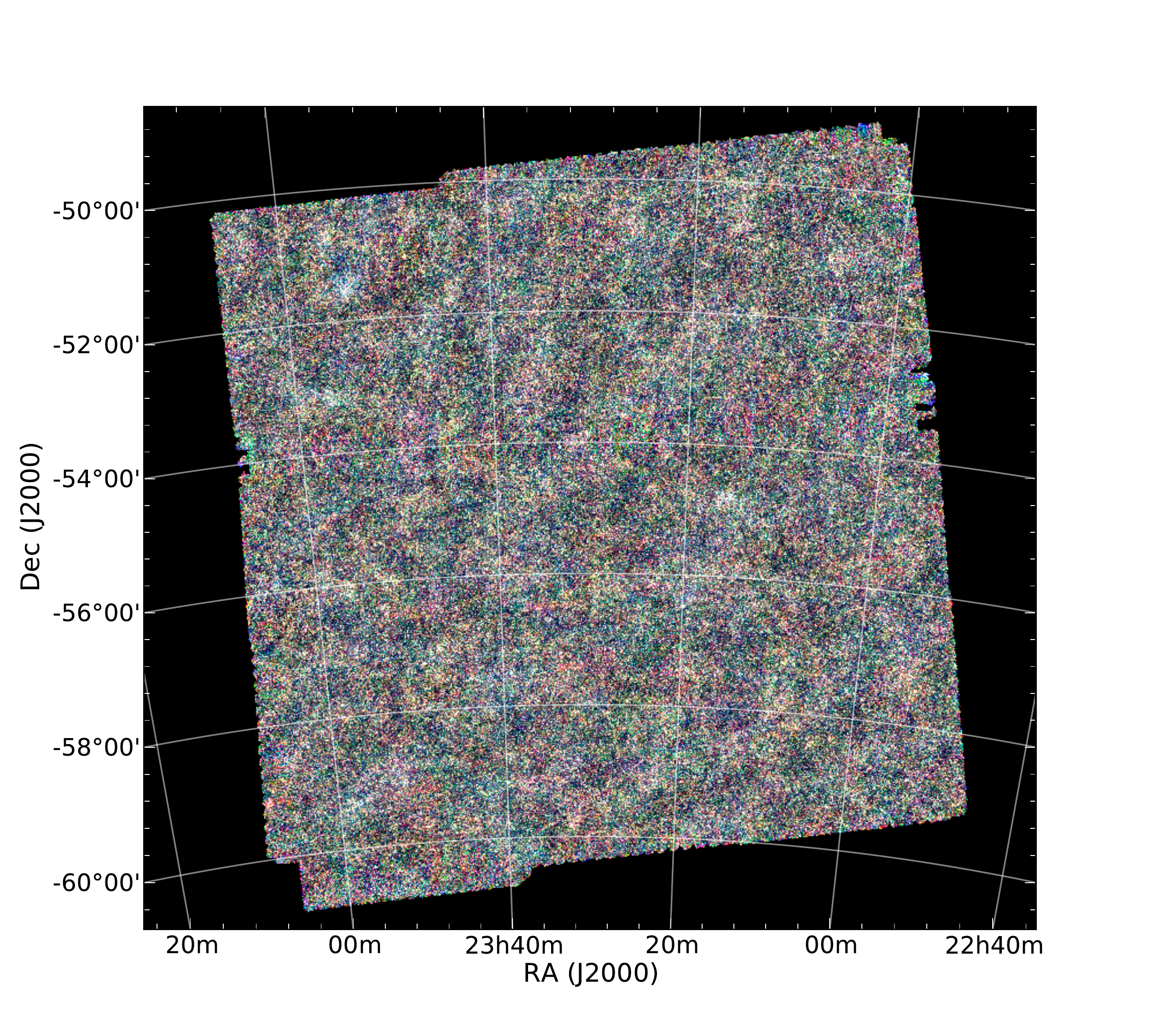}
\caption{SPT 150 GHz (2mm) temperature map (left) and \Herschel/\spire maps (right)
used for this analysis.
For display purposes only the inner 
$\sim 60\%$ of the SPT temperature map that was used to construct the lensing map is shown. 
In the right panel, (red, green, blue) correspond to (500, 350, 250)\rmicron.
}
\label{fig:maps}
\end{figure*}

A CMB map is generated as outlined in \citet{story12}. 
In addition, to avoid apodization
effects at the edges of the field when constructing the lensing map, 
data from surrounding fields are used to make
a single larger CMB map 17.1 degrees 
on a side. This map extends well beyond the region 
covered by \Herschel data. The input CMB map is shown in 
the left panel of Figure \ref{fig:maps}. Adjacent fields are combined
using inverse-variance weights in overlapping regions, and there is no
evidence for any discontinuities at the boundaries. Point sources
and massive galaxy clusters are removed using a Wiener-interpolation
algorithm \citep{vanengelen12}.

Simulated CMB maps are obtained by coadding 
simulated signal and noise realizations for each individual SPT field. 
The simulated maps are made with known input gravitational potentials, and
simulated signal maps are generated using timestream-based simulations, as in 
\citet{story12}. Noise realizations
are obtained directly from the observations, by taking randomized combinations of
the data which remove all sky signal, as detailed in
\citet{vanengelen12}. 

The analysis procedure is applied to both the real and simulated SPT maps.
Gravitational convergence maps are generated as outlined in 
\citet{vanengelen12}, using the quadratic estimator method \citep{hu01b,hu02a}.
This method entails constructing a gradient-filtered map and an inverse-variance
weighted map (i.e., two different filterings of the same CMB field), 
multiplying them together and taking a divergence. The resulting product
can be shown to be an estimator for the map of the gravitational potential.
The effective transfer function due to the SPT filtering was constructed by 
cross-correlating the derived lensing potential of the simulated maps
with the lensing potential maps used to generate those simulations.

Foreground contamination of the lensing convergence maps is expected to be small:
\citet{vanengelen12} found that residual contamination of the lensing convergence
map from point sources and galaxy clusters is expected to be at the level of a few $\%$.
The sign of this effect is expected to be negative, such that foreground contamination
acts to reduce the observed cross-correlation.

The resulting lensing convergence map is shown as contours in Figure \ref{fig:smoothed_maps}. Features
can be seen with significances exceeding $4\sigma$.

\section{\Herschel/\spire Maps}
\label{sec:spire}

Submillimeter maps at 500, 350, and 250\rmicron\ are created
using observations with the \spire instrument \citep{griffin03} aboard the {\em Herschel Space Observatory} \citep{pilbratt10} obtained under
an OT1 program (PI:Carlstrom).
Observations were made in \spire fast-scan mode ($60\, {\rm arcsec \, s^{-1}}$) and consisted of two sets of orthogonal scans covering $\sim~90\, \rm deg^2$.  
The observing strategy was chosen to optimize sensitivity to large-scale signal and provide redundancy for measuring the auto-frequency power spectrum of background fluctuations.     

Maps are made with {\sc smap}, an iterative mapmaker designed to optimally separate large-scale noise from signal;   
the mapmaking algorithm is described in detail in \citet{levenson10} and updated in \citet{viero12}.   
 To estimate the transfer function we
 use the same map-making process on mock SPIRE data.  For both real
 and mock data we make maps with 10 iterations; 
 we have checked that the maps are adequately converged at this point.
Additionally, time-ordered data (TODs) are divided into two halves and unique \lq\lq jack-knife\rq\rq\ map-pairs are made.  
To avoid having to reproject or regrid the \Herschel/\spire maps, 
we make them using the Lambert azimuthal equal-area projection
(also known as zenithal equal area, ZEA), 
with astrometry identical to that of the SPT map, and with 30\arcsec\ pixels.  

The maps have rms instrument noise levels (per 30\arcsec\ pixel) of 14, 10, and 7\,mJy, 
while the instrument effective point-spread functions 
are 36.6, 25.2, and 18.1 \arcsec\ full widths at half maxima (FWHM) at 500, 350, and 250\rmicron, respectively,
The 30\arcsec\ pixelization of the maps reduce the resolution substantially on small scales,  
but pixelization and instrument noise effects are not important on the scales of interest for this study.
The last step, following \citet{viero12} is to convert the maps from native units of $\rm Jy\, beam^{-1}$ to $\rm Jy\, sr^{-1}$, which is done by dividing them by the effective beam areas, 3.688, 1.730, and $1.053\times 10^{-8}$\,steradians. Color corrections from a flat-spectrum point-source calibration have a negligible effect.
The absolute calibration is accurate to 7\%, an uncertainty that is small compared to our statistical 
precision.

\section{Results and Analysis}

\begin{figure*}
\centerline{\includegraphics[width=6in]{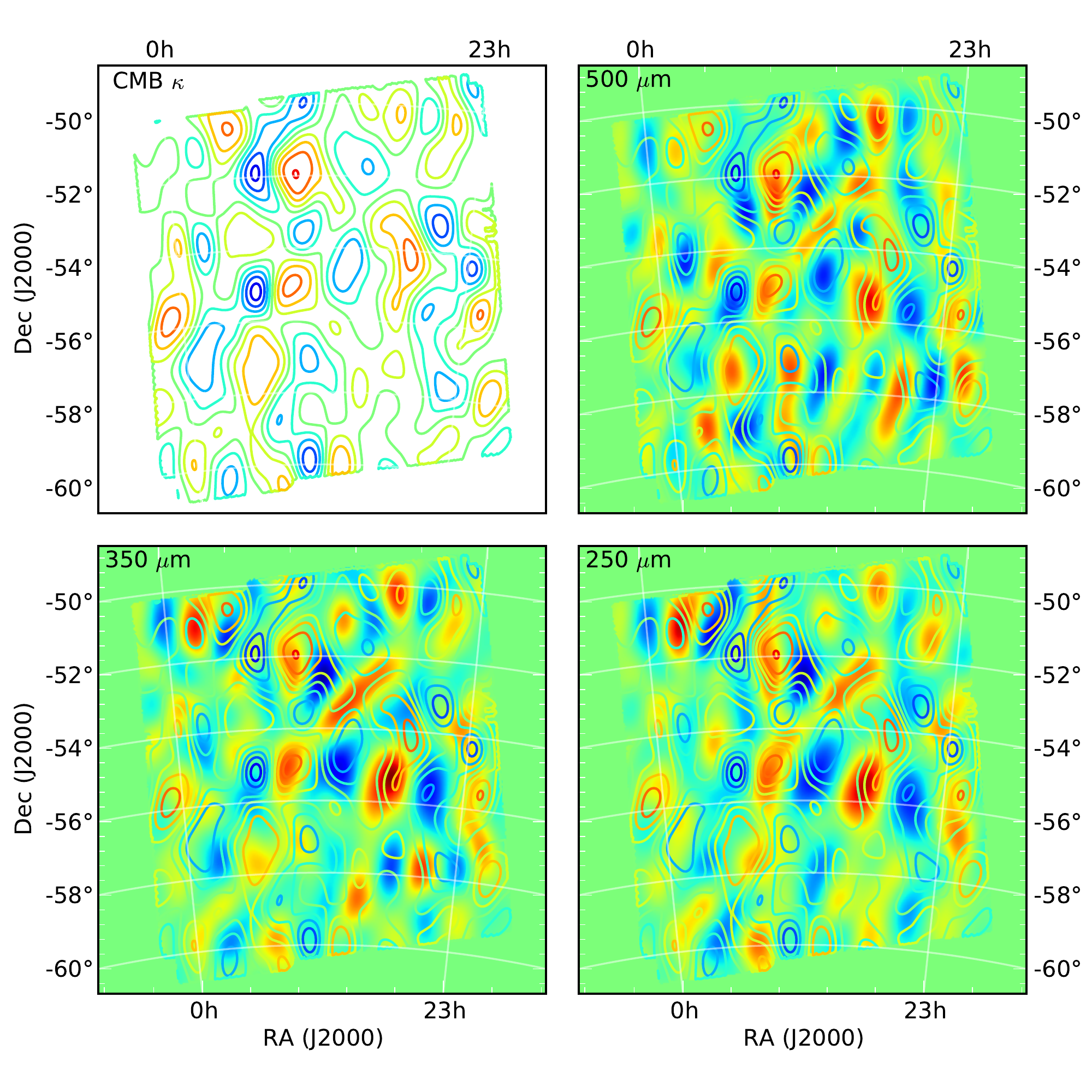}}
\caption{Map of the CMB lensing convergence measured with SPT data (contours in all panels) and overlaid
on maps of the 500, 350, 250 
\rmicron\ 
\Herschel/\spire\ data (top right, bottom left, bottom right, respectively). 
All maps have been filtered to only show scales in the 
lensing map that are expected
to have typical signal to noise of at least 0.5, which suppresses all
features on scales smaller than $\sim 0.5^{\circ}$. 
All maps have been masked by the \spire coverage. Lensing contours are spaced
by 1$\sigma$ of noise. Red (blue) indicates regions of increased (decreased) mass or flux.
\vskip0.33in}
\label{fig:smoothed_maps}
\end{figure*}

In Figure \ref{fig:smoothed_maps}, we present convergence and submillimeter-wavelengths maps 
filtered to emphasize modes in the lensing map that
have significant ($>$0.5) signal-to-noise, allowing a by-eye comparison of the structure. 
Modes with $L < 100$ (scales larger than $2^\circ$)
have been filtered to remove scales where the timestream filtering of the submm-wave
maps becomes substantial.
The SPT temperature
map has spatially anisotropic noise
\citep{schaffer11}, which ultimately leads to anisotropic noise in 
the lensing map \citep{vanengelen12}. This leads to a tendency for
modes to be better measured when they have more horizontal structure
than vertical structure.

Due to the imperfect redshift overlap, the
lensing map can have features that are not in the submm maps; in particular,
high-redshift structure ($z \gtrsim 3$) will appear relatively stronger in the lensing map, while structure
below $z \sim 0.5$ will be strongly suppressed in the lensing map as compared to the
submm map. 
The submm maps are extremely well-correlated with each
other, while the lensing
map has several features that are not well-matched in any of the submm 
maps. 
Nonetheless, there are many features in common between the
maps.

To compare these maps quantitatively, we use cross-power spectra, as in \citet{bleem12b}. 
Uncertainties are obtained by cross-correlating each submm map with
lensing mass maps obtained from simulated SPT maps. We use the rms in 
cross-power simulated amplitudes as the rms uncertainty and assume
a Gaussian error distribution. Cross powers are reported in Table
\ref{table:powers}, and are shown in Figure \ref{fig:cross_spectra}. 

\begin{table}[b]
\caption{CMB convergence-\spire Cross power spectrum}
\begin{tabular}{r r r r}
$L$ &  $C_L^{500}$    & $C_L^{350}$  & $ C_L^{250}$  \\
  &  (mJy/sr)   &  (mJy/sr) &  (mJy/sr) \\
\hline
150 &  $38 \pm 18 $ & $101\pm 29$  & $134\pm 58$ \\ 
250 &  $49 \pm 21 $ & $65\pm 30$  & $74\pm 45$ \\ 
350 &  $10 \pm 11 $ & $18\pm 15$  & $26\pm 21$ \\ 
450 &  $19.0 \pm 6.2 $ & $38\pm 10$  & $52\pm 13$ \\ 
550 &  $8.5 \pm 6.4 $ & $15\pm 10$  & $8\pm 12$ \\ 
650 &  $13.7 \pm 4.6 $ & $16.0\pm 7.8$  & $16.9\pm 8.8$ \\ 
750 &  $13.3 \pm 4.2 $ & $17.5\pm 5.6$  & $14.3 \pm 8.5$ \\ 
850 &  $4.9 \pm 3.2$ & $3.5\pm 5.4$  & $13.5\pm 7.0$ \\ 
950 &  $6.1 \pm 1.9$ & $9.5\pm 3.5$  & $7.7\pm 5.0$ \\ 
1050 &  $6.9 \pm 1.9$ & $8.1\pm 3.2$  & $2.1\pm 4.6$ \\ 
1150 &  $1.2\pm 1.5$ & $2.9\pm 2.4$  & $0.7\pm 3.6$ \\ 
1250 &  $5.7\pm 2.2$ & $9.2\pm 3.1$  & $11.6\pm 4.2$ \\ 
1350 &  $2.3\pm 1.5$ & $5.5\pm 2.6$  & $6.1\pm 3.5$ \\ 
1450 &  $4.2\pm 1.8$ & $5.1\pm 3.1$  & $3.8\pm 4.1$ \\ 
1550 &  $2.2\pm 1.6$ & $2.4\pm 2.9$  & $3.1 \pm 3.3$ \\ 

\hline
\hline
\end{tabular}
\label{table:powers}
\end{table}

The signal-to-noise ratio in the cross correlation
is substantial: at 500, 350, and 250 \rmicron\ the model with no cross-correlation is strongly
disfavored relative to the best-fit lensing amplitude, with $\chi^2$ differences of  
79, 69, and 45, respectively.  Lensing is detected in every power spectrum
bin.

\section{Theoretical Model}

As a cross-check on the shape and amplitude of these spectra, we adopt the 
simple constant bias model used in \citet{bleem12b}, using the non-linear 
power spectrum at each redshift: 

\begin{equation}
C_L^{\kappa \mathrm{I}} = b \int dz\, {d\chi \over dz} {1\over\chi^2} W^\kappa(\chi) 
W^\mathrm{I}(\chi) P_\mathrm{DM}\left(k = {L \over \chi}, z\right),
\label{eq:clcross}
\end{equation}
where $W^\kappa(\chi)$ gives the redshift weighting of the mass map
and $W^\mathrm{I}(\chi)$ is proportional to the line of sight distribution
of the intensity $dI/d\chi$ \citep{bleem12b, song03}.  
The non-linear power spectrum of the
dark matter, $P_\mathrm{DM}$, is calculated using 
CAMB and Halofit, assuming the best-fit WMAP9+SPT cosmological parameters
for a flat $\Lambda$CDM cosmology \citep{story12}.

\begin{figure*}
\centerline{
\includegraphics[width=2.63in]{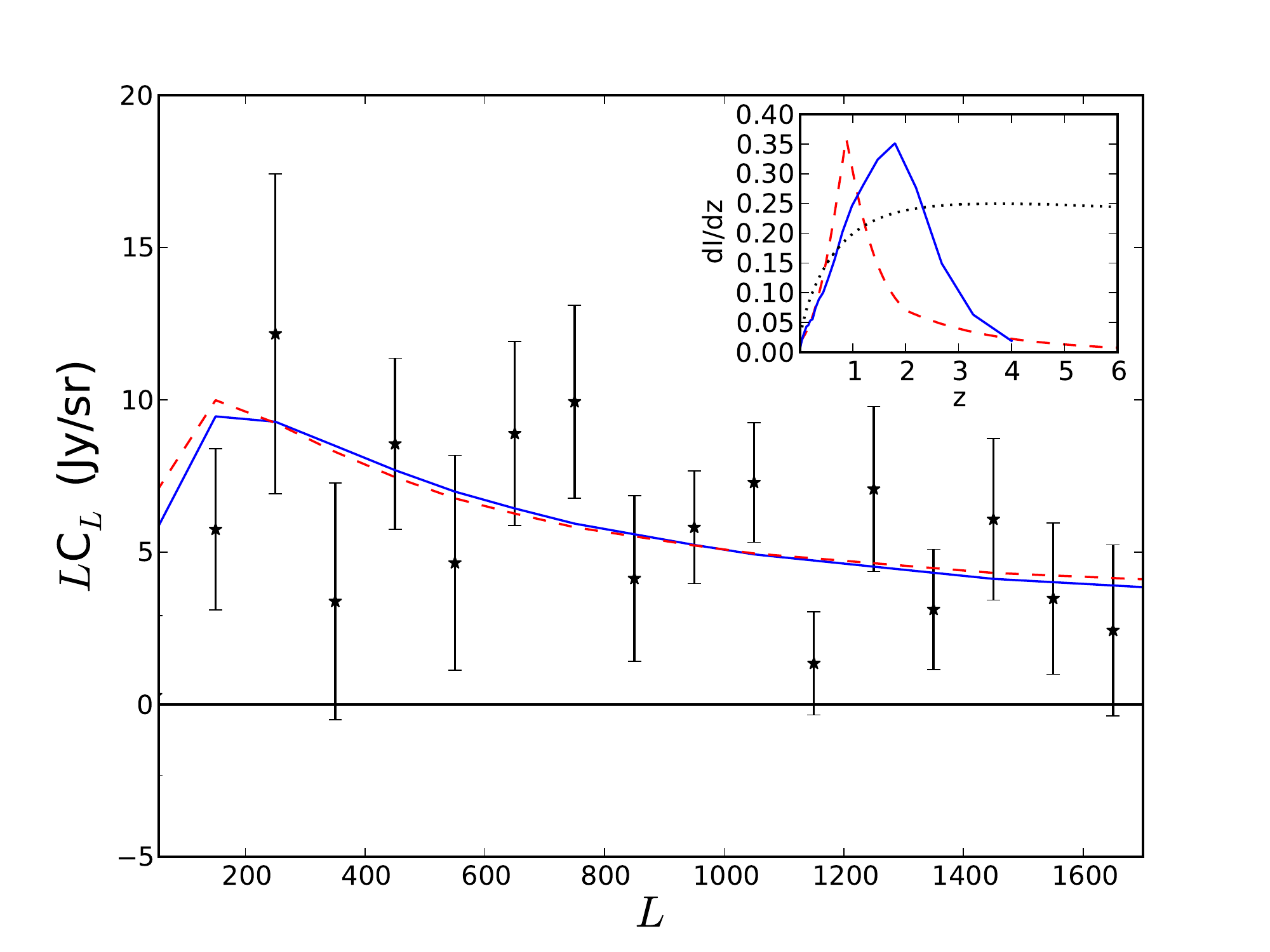}
\hskip-0.29in \includegraphics[width=2.63in]{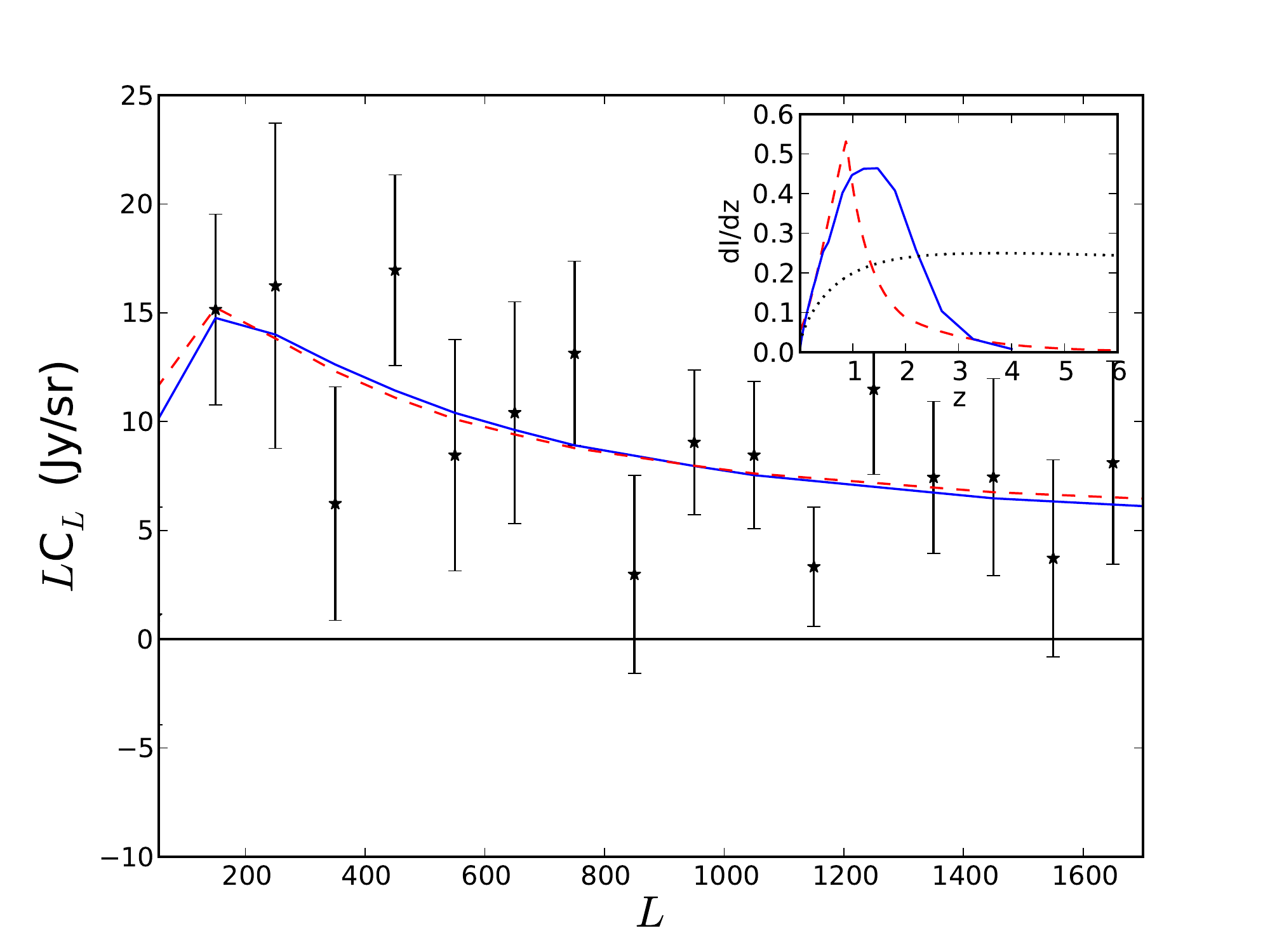}
\hskip-0.29in \includegraphics[width=2.63in]{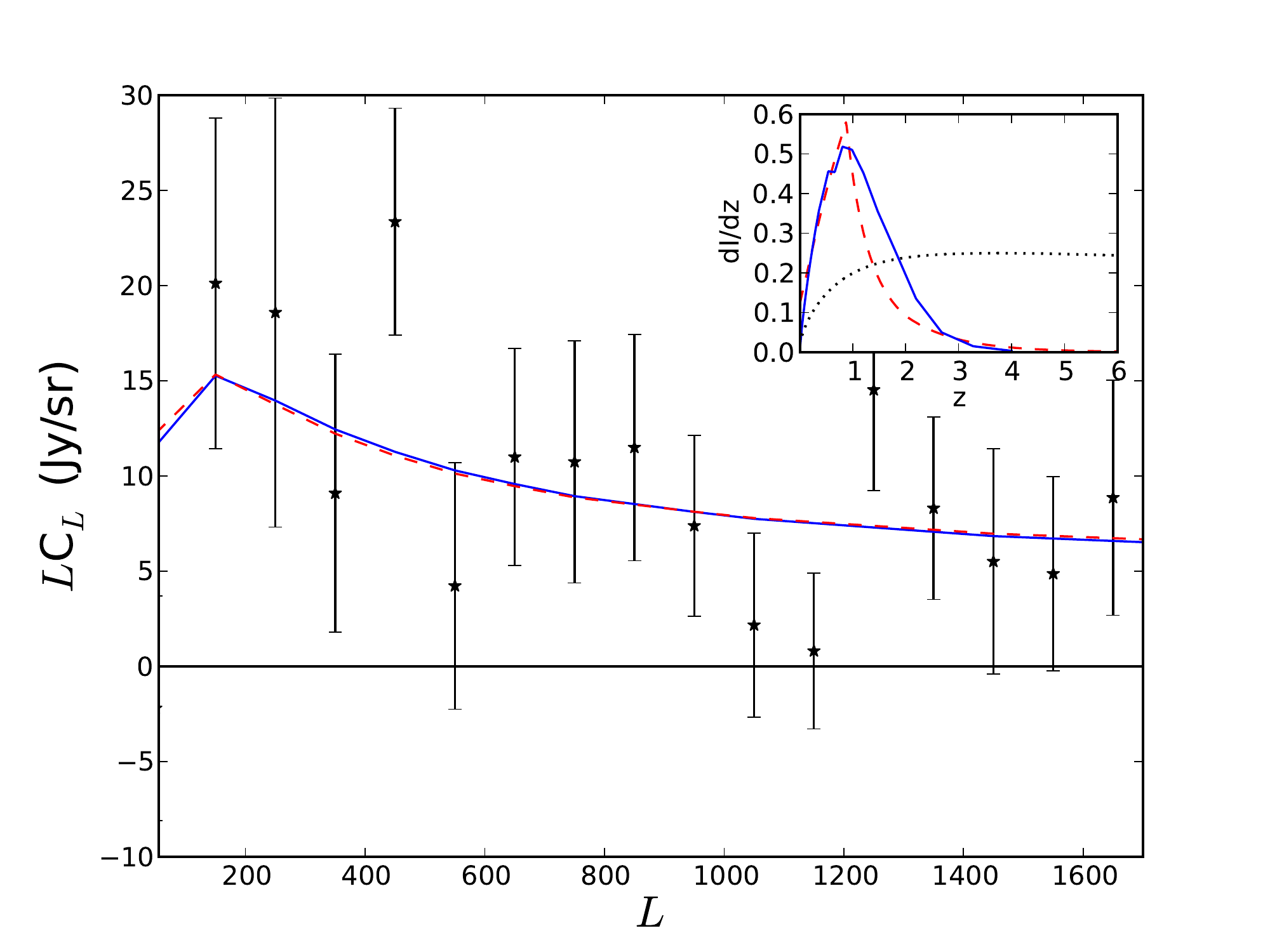}
}
\caption{Cross-spectrum of lensing map and submm maps: left to right
show 500, 350, 250 \rmicron. Overplotted
are best-fit constant bias models for two different
redshift distributions for the submm intensity (with $dI/dz$ shown in inset in units of MJy/sr),
with red (dashed) showing the model of \citet{bethermin11} and blue showing
the model of \citet{viero12}. Also shown in the inset, in arbitrary units, is the
weighting of the CMB lensing kernel as a function of redshift (black dotted). }
\label{fig:cross_spectra}
\end{figure*}

The redshift distribution of contributions to the submm background
has been extensively studied in recent years, 
and there exist substantial disagreements between authors.
We adopt two recent determinations,
presented in \citet{bethermin11} and \citet{viero12} 
that roughly bracket expectations, to predict the 
cross-correlation signal.  
To derive this signal, 
we assume that the submm light traces the non-linear dark matter
density field at every redshift, with a single amplitude, the
bias $b$, that we fit to the data. The cross-correlation will be most
sensitive to redshifts $z\sim 0.5-3$, with lower $z$ a poor match to
CMB lensing, and higher $z$ not having substantial submm emission. 
As seen in the insets of Figure \ref{fig:cross_spectra}, the 500 \rmicron\
emission is expected to have broader overlap with the CMB lensing
kernel, and should therefore show a stronger correlation.

Fits are performed using points between $L=100$ and $L=1600$, 
as done in previous SPT lensing studies.
The best-fit bias parameters for each observing wavelength and redshift 
distribution choice are shown in Table \ref{table:biases}, with best-fit
bias parameters depending on which redshift
distribution is assumed. 
For the \citet{bethermin11} model we find 
$b\sim 1.8\pm 0.3$ while the \citet{viero12} model for the CIB intensity
gives $b\sim 1.3 \pm 0.2$. The uncertainties reflect statistical uncertainties only,
and the large difference between the two models indicates that systematic uncertainties
are substantial.
The difference in bias factors is largely due
to the different integrated mean intensities in the two models; for example,
at 500 \rmicron\ the two models predict mean intensities that differ by a factor
of 1.5, while the derived bias factors differ by a factor of 1.4. This difference
in the mean intensity is larger than the $\sim 25\%$ uncertainty in the FIRAS measurements
\citep{fixsen98}; the mean intensity in the \citet{viero12} model is more than $2\sigma$ 
higher than that measured by FIRAS at 500 \rmicron.

This simple model provides a very good fit, 
with $\chi^2=12.6$ or 12.7 for 14 degrees of
freedom at 500 \rmicron, depending on the assumed redshift distribution
of the submm background.  Despite the qualitative difference
in the two redshift distributions apparent in the insets of Figure
\ref{fig:cross_spectra}, good fits are obtained for both models, although 
a different normalization is preferred by each. This arises because
most of the power is coming from the non-linear regime, where a
power-law is a remarkably good fit to the clustering power spectra
\citep{addison12}. 
As the cross-spectrum is a superposition of similar power-laws from different
epochs, the detailed redshift distribution 
does not affect the shape of the cross-spectrum.

\begin{table}
\caption{Fits to constant bias model}
\begin{tabular}{c c c}
Wavelength  &  Bias (V12)  & Bias (B11)  \\
\hline
500 \rmicron &  $1.29 \pm 0.16 $ (12.6) & $1.80 \pm 0.22$ (12.7) \\
350 \rmicron &  $1.35 \pm 0.17$ (9.7) &  $1.82 \pm 0.24$ (9.9)    \\
250 \rmicron &  $1.34 \pm 0.23$ (11.8) &  $1.56 \pm 0.27$ (12.0)    \\
\hline
\end{tabular}
\begin{tablenotes}[para]
Cross-spectrum best-fit amplitudes to constant bias model
for Viero et al (2012) (V12) and Bethermin et al. (2011) (B11) redshift
distributions, $\chi^2$ of fit shown in parentheses. Quoted uncertainties only
include statistical uncertainty.
\end{tablenotes}
\label{table:biases}
\end{table}

The bias factors at infrared and
submm wavelengths have been measured using
both source catalogs and auto power spectra of
the diffuse backgrounds, as reviewed recently
for the cosmic infrared background in \citet{penin12}. The inferred
bias values depend on the assumed redshift distribution and intensity of
the background, and the bias value that we measure is the clustering amplitude
relative to the non-linear matter power spectrum, rather than either
the linear matter power spectrum or a halo model, so a direct comparison
is difficult. 
Using BLAST data
at 500, 350, and 250 \rmicron\ \citep{viero09} and intensity estimates from \citet{lagache04},
typical bias factors of $2.2 \pm 0.2$ were found \citep{penin12}. 
\citet{amblard11} find slightly higher 
bias values using a halo model and fitting internally for the
intensity as a function of redshift. The bias values found here are
somewhat lower, but could be explained by differences in the assumed
mean intensities and their redshift distributions. 

Some studies of dusty sources at high redshift have led to substantially higher
bias factors: \citet{brodwin08} found that $z\sim 2$ dusty, obscured galaxies selected
in the optical/IR had bias factors $b\sim 3-5$, while \citet{hickox12} used sources
selected at 870 \rmicron\ to estimate $b\sim 3$.

For comparison with lower redshift galaxy samples, 
recent results from SDSS-III find
bias factors of $\sim 2$ for the massive galaxies 
(halo masses $\sim 5 \times 10^{13} h^{-1} M_\odot$) being targeted
for baryon acoustic oscillation studies at $z\sim 0.3$ \citep{parejko13},
	while bias estimates based on the SDSS main galaxy sample
	\citep{mcbride11} find $b=1-1.2$ for typical luminosity ($L_*$). 
	This suggests that the typical contributors to the submm background
	could be the higher redshift precursors  to (or at least have the
	same mean bias as) galaxies 
	that are intermediate in mass between these two
	samples. 

In work that is closely related to the current work, \citet{hildebrandt13} 
cross-correlated gravitational lensing of Lyman-break galaxies with a catalog
of sources detected at 250 \rmicron, and inferred typical masses of 
$1.5 \times 10^{13} M_\odot$ for these galaxies.  

\section{Discussion and Conclusions}

We have shown that large-scale structure traced by submm sources is
well-correlated with a CMB lensing convergence map.  The cross-correlation is
highly significant at 500, 350, and 250 \rmicron, corresponding to
detection significances of 8.9, 8.3, and 6.7 $\sigma$, respectively. 

The cross-correlation between the lensing convergence map and
each submm map is well fit by a simple constant bias model, 
with bias factors of $b=1.3-1.8$, depending on the assumed
redshift distribution for the submm intensity.
The lower bias factors are found for an assumed intensity distribution 
with more flux coming from higher redshifts.

There are several ways to extend the utility of the lensing
convergence-SPIRE
 cross-power spectra presented here.  For example, combining them with the
 cross-power and auto-power spectra among the three SPIRE bands will
 probe the redshift distribution of the contributing sources and the
 correspondence between submm flux and the underlying dark matter
 distribution.

This technique is highly complementary to studies of the auto- and
cross-correlations of submm background maps. While convergence maps have more noise (at current
CMB map noise levels), concerns about Galactic cirrus or separating shot noise are
greatly reduced, making cross correlation with CMB lensing an extremely robust
probe of clustering with a promising future. 

With the release of {\em Planck} maps covering a broad range of CIB wavelengths with well-matched
angular resolution, it will be possible to perform a similar analysis over the entire 2500 square degree
SPT survey area, while the coming Dark Energy Survey (DES) will also have nearly complete overlap with this area. 
DES will have both galaxy catalogs and cosmic shear maps with some resolution in the 
line of sight direction. In combination with the SPT CMB lensing convergence this will enable
3D mass maps of the universe extending to $z\sim~1100$.

\begin{acknowledgements}
The SPT is supported by the National Science Foundation through grant ANT-0638937, with partial support provided by
 NSF grant PHY-1125897, the Kavli Foundation, and the Gordon and Betty Moore Foundation.
 The McGill group acknowledges funding from the National Sciences and Engineering Research Council of Canada, Canada
  Research Chairs program, and the Canadian Institute for Advanced Research.
  Work at Harvard is supported by grant AST-1009012.
  S.~Bhattacharya acknowledges support from NSF grant AST- 1009811,
  R.~Keisler from NASA Hubble Fellowship grant HF-51275.01,
  B.~Benson from a KICP Fellowship,
  M.~Dobbs from an Alfred P. Sloan Research Fellowship,
  and O.~Zahn from a BCCP fellowship.

\end{acknowledgements}

\bibliography{../../BIBTEX/spt}

\end{document}